# High Ferromagnetic Transition Temperature (~172K) in Mn $\delta$-doped GaAs with *p*-type Selective Doping

**Ahsan M. Nazmul [a), b)], S. Sugahara [a), b)], and M. Tanaka [a), b)]**

(a) Department of Electronic Engineering, University of Tokyo,
7-3-1 Hongo, Bunkyo-ku, Tokyo 113-8654, Japan.

(b) PRESTO, Japan Science & Technology Corporation,
4-1-8 Honcho, Kawaguchi, Saitama 332-0012, Japan.

**Abstract.** We have found high ferromagnetic transition temperature in Mn δ-doped GaAs-based heterostructures grown on GaAs(001) substrates by molecular beam epitaxy. A 0.3 ML Mn δ-doped GaAs samples showed high resistivity at low temperature and did not show a ferromagnetic behavior. However, in a selectively doped heterostructure (Mn δ-doped GaAs / Be-doped AlGaAs), where holes were supplied from the Be-doped AlGaAs layer, clear ferromagnetic order was observed. The ferromagnetic transition temperature $T_C$ of the selectively doped heterostructure was as high as 172K with suitable low-temperature (LT) annealing treatment.

## 1. Introduction

Recently, there are growing efforts to manipulate spin-related functions in semiconductor-based materials to develop novel spin-electronic (or "spintronic") physics and devices. For this purpose, various new materials and structures have been developed based on III-V semiconductors. Among them, (1) ferromagnet/semiconductor hetetrostrcutures such as MnAs/GaAs [1] and (2) magnetic semiconductor alloys such as InMnAs and GaMnAs [2-3] have been the focus of studies. Although both material systems are extensively investigated, in the former system special techniques are required to grow multilayer heterostructures with abrupt interfaces [4] and much higher ferromagnetic transition temperature $T_C$ is needed for the latter system (for example, the highest $T_c$ reported to date is 110K for (GaMn)As [5]).

We investigate a new type of spin-electronic structure; Mn δ-doped GaAs-based heterostructure. We aim to broaden the degree of freedom in manipulating spin-related properties and functions by utilizing the inherent advantages of the δ-doping technique; that is, locally high doping concentration, and also high carrier concentration, as are reported in Be and Si-δ-doped GaAs layers [6]. Another prospective feature of the Mn δ-doping is easy fabrication of multiplayer heterostructures using Mn δ-doped GaAs layers with excellent interfaces, which are also essentially important for spintronic devices that are particularly sensitive to the quality of interfaces. Structural analysis using secondary ion mass spectroscopy (SIMS) and transmission electron microscopy (TEM) characterizations of low temperature (LT) molecular beam epitaxy (MBE) grown

Mn δ-doped GaAs layers in our earlier reports revealed that most of the Mn atoms are abruptly confined within a width of 2 monolayers (ML), when the nominal thickness $\theta_{Mn}$ of Mn is less than 1 ML [7,8]. The Mn doping profiles retained abruptness even at elevated growth temperature $T_s$ up to 400°C. Here, $\theta_{Mn}$ = 1 ML corresponds to a sheet Mn concentration of $6.3\times10^{14}$ cm$^{-2}$, assuming that Mn atoms are substituted for the Ga sublattice in the zinc-blende GaAs with the lattice constant of 0.565 nm. The TEM and X-ray standing wave method studies showed that the local structure around Mn maintains zinc-blende type crystal structure, indicating that Mn atoms are substituted for the Ga sublattice in our Mn δ-doped GaAs [9]. Although it was possible to incorporate high Mn dopant concentration in the Mn δ-doped GaAs layers, the hole to dopant concentration ratio $p/\theta_{Mn}$ was very low and was not enough to realize ferromagnetic ordering [8]. In this paper, we focus on a selectively doped heterostructure (Mn δ-doped GaAs / Be-doped AlGaAs), where holes were intentionally supplied from the Be-doped AlGaAs layer to the Mn δ-doped GaAs layer reflecting the semiconductor band-engineering.

## 2. Structure and MBE growth

We have grown 0.3ML Mn δ-doped GaAs / Be-doped *p*-AlGaAs heterostructures using LT-MBE, whose structure is shown in Fig. 1(a). The structure of the selective doping approach in the sample resembles an inverted high electron mobility transistor (I-HEMT). A 200 nm-thick undoped GaAs buffer, a 300 nm-thick $Al_{0.3}Ga_{0.7}As$, a 30 nm-thick Be-doped $Al_{0.3}Ga_{0.7}As$ (Be concentration was $1.8\times10^{18}$ cm$^{-3}$), and a $d_s$ nm-thick GaAs separation layer were successively grown at $T_s$ = 600°C on a semi-insulating (SI)-GaAs(001) substrate. Then, a Mn δ-doped GaAs layer with Mn concentration $\theta_{Mn}$ = 0.3ML and a subsequent 20 nm-thick undoped GaAs cap layer were grown at $T_s$ = 400°C and 300°C. The thickness ($d_s$) of the undoped-GaAs seperation layer was a measure to control the interaction between the Mn δ-doped GaAs layer and the 2-dimensional hole gas (2DHG) formed at the GaAs/AlGaAs interface. Successful epitaxial growth of the heterostructure was confirmed by *in-situ* RHEED observations. Figure 1 (b) shows the cross-sectional high-resolution transmission electron microscopy (TEM) image of a 1.0 ML Mn δ-doped GaAs layer grown at $T_s$ = 300°C. Mn dopants are abruptly localized in the zinc-blende structure within 2~3 ML-width without formation of dislocations or clusters, indicating the excellent crystalline property of Mn δ-doped GaAs layers.

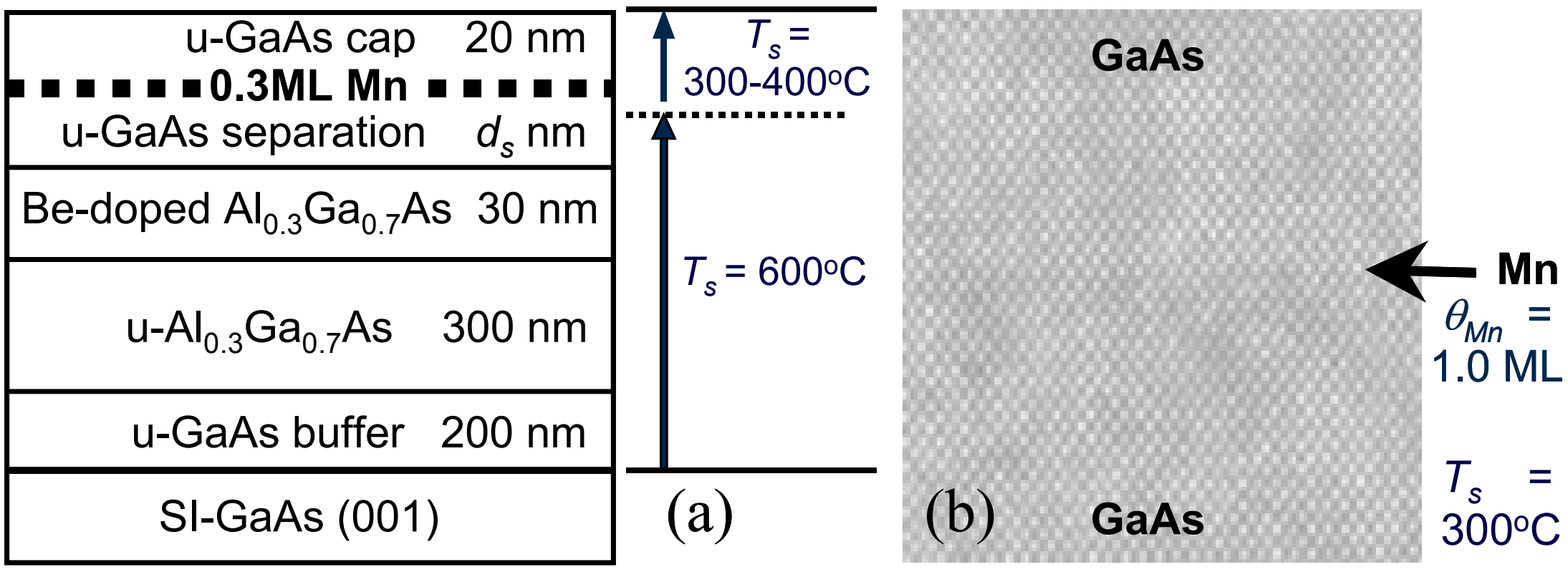

**Figure 1**. (a) Sample structure of a Mn δ-doped layer with a selectively doped I-HEMT structure. Holes are supplied into the Mn δ-doped GaAs layer from the Be-doped $Al_{0.3}Ga_{0.7}As$ layer. The GaAs separation layer thickness $d_s$ was 0 ~ 10 nm. (b) High-resolution TEM lattice image of the Mn δ-doped GaAs layer at $\theta_{Mn}$ = 1.0 ML grown at $T_s$ = 300°C.

## 3. Transport and magnetic properties

We have investigated the magneto-transport properties of the samples in Hall-bar geometry. Hall measurements of 0.3 ML Mn δ-doped GaAs layers grown at 400°C *without* and *with* the I-HEMT structure ($d_s$ = 3 nm) are shown in Fig. 2 (a) and (b), respectively. The hysteresis in the Hall loop of the sample *with* I-HEMT clearly indicates ferromagnetic order, while the ferromagnetic hysteresis is absent in the sample *without* I-HEMT. The origin of the negative slope in Fig. 2 (b) is the influence of the negative magneto-resistance, which is proportional to the Hall resistivity when skew scattering of the holes is dominant. Details of the expressions are described later. The temperature dependence of sheet resistance ($R_{sheet} - T$) of the samples *without* and *with* I-HEMT is plotted in Fig. 2 (c). The sample *without* I-HEMT shows an insulating behaviour. In contrast, the sample *with* I-HEMT shows a metallic behaviour, and a local maximum of the $R_{sheet} - T$ trace suggests that $T_C$ is as high as 70K. This value of $T_C$ was confirmed by measuring Hall loops where hysteresis remained open till 60 ~ 70K and closed at 80K. This indicates that the combination of localized magnetic ions in a δ-doped profile and *p*-type selective doping induced ferromagnetism in GaAs.

The ferromagnetic order of the samples was found strongly dependent on the GaAs separation layer thickness $d_s$. As shown in Fig. 3 (a), the local maximum temperature of the bump in the $R_{sheet} - T$ trace, roughly corresponding to the Curie temperature $T_C$, was 45K at $d_s$ = 0 nm and 70K at $d_s$ = 3 nm. The Hall loops showed clear ferromagnetic hysteresis at $d_s$ = 0 and 3 nm below $T_C$. With further increase in $d_s$ to 5 and 10nm, the bump disappeared. Hysteresis in loops was not observed at $d_s$ = 5 and 10 nm indicating the absence of ferromagnetic order. The dependence of the ferromagnetic order on $d_s$ is explained using the band diagram of the structure as shown in Fig. 3 (b). Here, *z* is the growth direction of the sample. Two dimensional hole gas (2DHG) was formed in the GaAs nearby the heterointerface. The hole concentration of the 2DHG in the I-HEMT without Mn δ-doped GaAs layer was $1.8\times10^{12}$ $cm^{-2}$ estimated by Hall measurements. We think that the degree of overlap of the wave function of the 2DHG and the Mn δ-doping profile directly affects the ferromagnetic ordering and $T_C$ of the structure. $T_C$ was highest (70K) at $d_s$ = 3 nm, probably because the overlap of the wave function of the 2DHG and the Mn δ-doping profile was maximum at $d_s$ = 3 nm. Further increase in $d_s$ resulted in the decrease in the overlap of the wave functions and weakened the ferromagnetic order. We further notice that the sheet resistance $R_{sheet}$ of the Mn δ-doped GaAs heterosturcures decreased with the increase of the separation layer thickness $d_s$ (see Fig. 3 (a)). We explain the phenomenon with a scattering mechanism. It is apparent that the GaAs separation layer thickness $d_s$ plays the role of controlling the degree of scattering of the 2DHG wave function by the Coulomb potential of ionized Mn dopants in the δ-doped layer. The increase in the $d_s$ will weaken the scattering, thus will increase the mobility of the 2DHG and eventually will decrease the sheet resistance $R_{sheet}$. This indicates that the ferromagnetic order in the I-HEMT structure can be intentionally controlled by the thickness of the separation layer.

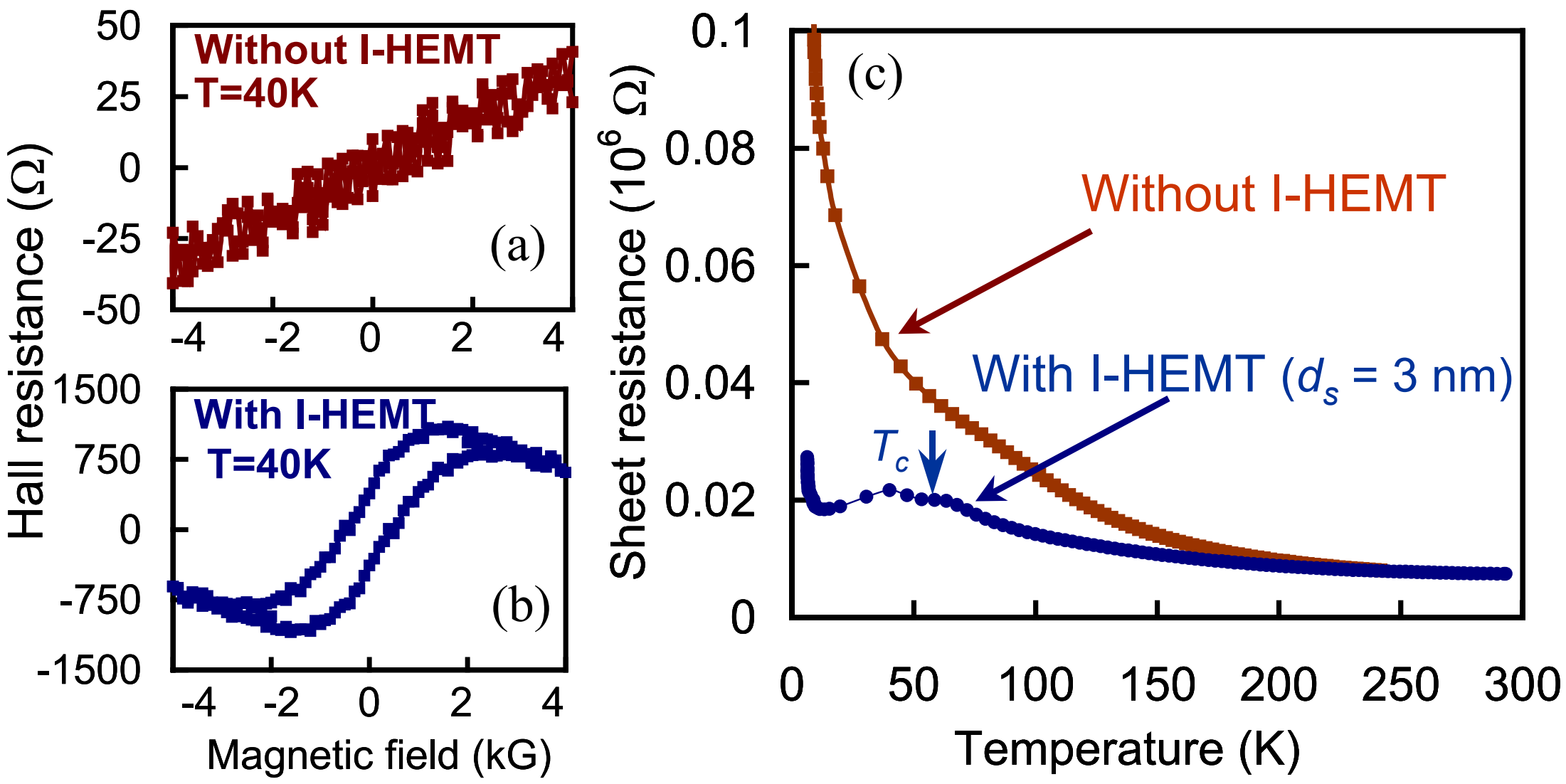

**Figure 2**. (a) and (b) show Hall loops of 0.3 ML Mn δ-doped GaAs layers grown at 400°C *without* and *with* I-HEMT ($d_s$ = 3 nm), respectively, indicating a ferromagnetic order in the I-HEMT structure at 40K; (c) $R_{sheet} - T$ traces of the samples *without* and *with* I-HEMT, respectively.

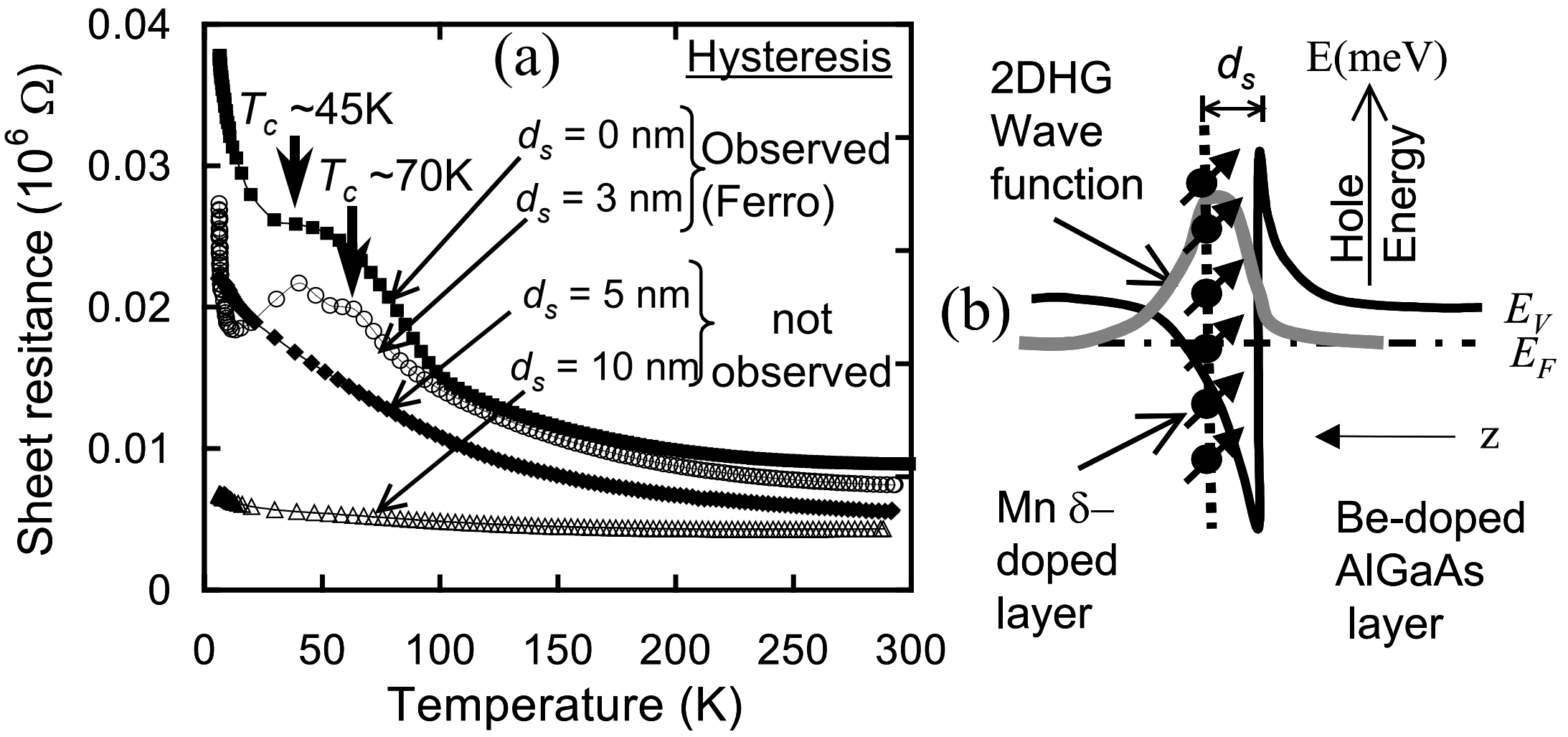

**Figure 3**. (a) $R_{sheet}$ – $T$ traces of 0.3 ML Mn δ-doped GaAs samples *with* I-HEMT for $d_s$ = 0, 3, 5 and 10 nm, respectively, grown at 400°C. The local maxima indicating $T_C$ increased from 45K to 70K when the $d_s$ changed from 0 to 3 nm, however, disappeared at $d_s$ = 5 and 10 nm. The ferromagnetic hysteresis in hall loops appear at $d_s$ = 0 and 3 nm, and disappeared at $d_s$ = 5 and 10 nm. (b) Schematic diagram of the band profile of the structure, the wave function of two-dimensional hole gas (2DHG), and Mn dopants. $E_V$ and $E_F$ correspond to the Fermi energy level and valence band top, respectively. $z$ is the growth direction.

## 4. Low temperature (LT) MBE growth and LT-annealing

In order to completely suppress the surface segregation of Mn in the Mn δ-doped GaAs layer, $T_s$ of the Mn δ-doped GaAs layer in the I-HEMT was lowered from 400°C (at which surface segregation of around 30% Mn dopants was observed in the SIMS depth profile) to 300°C (at which no segregation of Mn was seen) [7]. Although the hole to dopant concentration ratio $p/\theta_{Mn}$ in a single Mn δ-doped GaAs layer at 300°C is only 0.001 comparing to 0.040 grown at 400°C, the suppression of Mn surface segregation at $T_s$ = 300°C will increase the local Mn concentration while holes are intentionally provided from a Be-doped AlGaAs layer. The sample structure of the 0.3ML Mn δ-doped GaAs ($T_s$ = 300°C) / Be-doped *p*-AlGaAs heterostructure examined here was similar to Fig. 1 (a) with $d_s$ = 0 nm.

### *4.1. Hall measurements*

Hall measurements were carried out in a Hall-bar geometry. It is notable that in the LT-MBE method, the growth temperature is far below the equilibrium growth condition. The transport and magnetic properties are therefore highly sensitive to the growth conditions (substrate temperature and As pressure) [10]. Kastsumoto et al reported that low temperature (LT) annealing treatment (at temperatures slightly above the growth temperature) was found unexpectedly effective to improve the transport and magnetic properties [11]. We also expect a similar effect for the Mn δ-doped GaAs grown layer grown at $T_s$ = 300°C in the I-HEMT structure. The LT-annealing treatment was carried out in a nitrogen atmosphere for 15 minutes with annealing temperature $T_a$ at 280, 300, 320 and 335°C, respectively. Fig. 4 shows the Hall loops of the as-grown and annealed samples. The hysteresis in the Hall loop in Fig. 4 (a) indicates that ferromagnetism was revealed at 100K in the as-grown sample. Remarkable improvement in the magnetic property is achieved in the annealed samples (Fig. 4 (b)-(d)). Especially, the hysteresis in the Hall loop in Fig. 4 (c) indicates that the ferromagnetic order is retained even at 170K for the sample annealed at $T_a$ = 300°C. Further increase in the annealing temperature above 300°C resulted in the fall of ferromagnetic transition temperature. Fig. 4 (d) shows ferromagnetic hysteresis in the

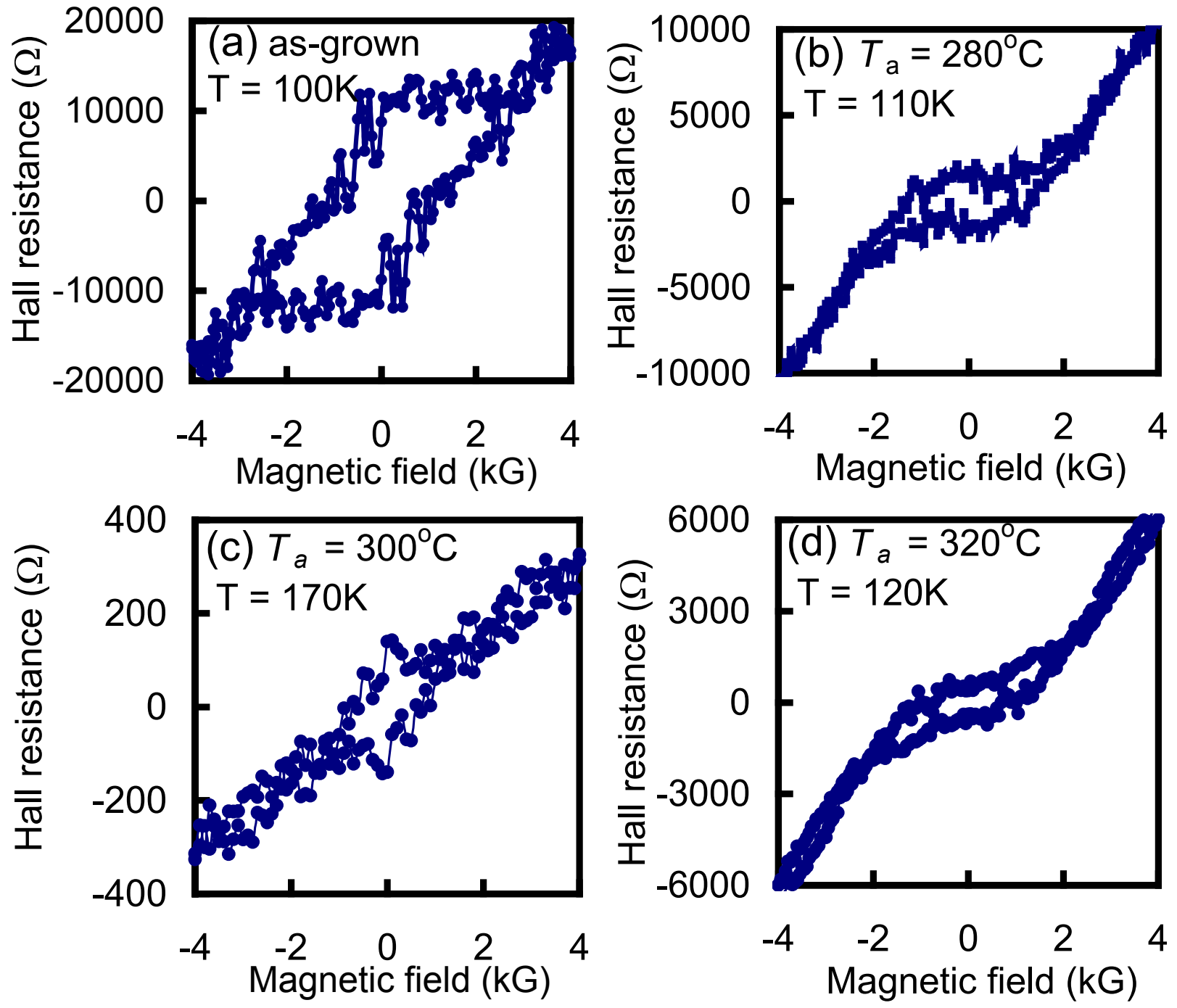

**Figure 4**. Ferromagnetic Hall hyteresis loops of a 0.3 ML Mn δ-doped GaAs layer *with* I-HEMT ($d_s$ = 0 nm) grown at 300°C; (a) as-grown sample measured at 100K, (b) annealed sample with $T_a$ = 280°C measured at 110K, (c) annealed sample with $T_a$ = 300°C measured at 170K, and (d) annealed sample with $T_a$ = 320°C measured at 120K.

Hall loop measured at 120K for the sample annealed at $T_a$ = 320°C. When the sample was annealed at $T_a$ = 335°C, the transport properties drastically deteriorated. No ferromagnetic hysteresis loops were detected in the capability of the measurement system.

Our experiments show that the sample grown at 300°C resulted in the increase of $T_C$ from 70K (grown at $T_s$ = 400°C) to 112K (as-grown), and 172K (LT-annealed). While the partially segregated Mn dopants at $T_s$ = 400°C do not take part in the ferromagnetic ordering, the more abrupt profile of Mn dopants with higher peak concentration grown at 300°C resulted in the increase of $T_C$ in the I-HEMT structure. This increase of $T_C$ was further enhanced by the LT-annealing treatment, which contributed to the reduction of point defects and thus compensation.

*4.2. Curie Weiss fitting to the transport properties*

In order to estimate the Curie temperature $T_C$ and the hole concentration of the sample by ruling out the anomalous Hall effect, we have performed the Curie Weiss fitting to the experimental trace of Hall resistivity $\rho_H$ (or Hall coefficient $R_H$) versus temperature $T$. In a magnetic material, there exists the anomalous Hall effect contribution to the Hall resistivity $\rho_H$, expressed as $\rho_H = R_O B + R_S M$ (Where $R_O$ is the ordinary Hall coefficient, $B$ is the magnetic field, $R_S$ is the anomalous Hall coefficient and $M$ is the magnetization of the sample) [12]. $R_S$ is proportional to the resistivity $\rho$ of the sample, $R_S = c\rho$ ($c$ is a proportional constant), when skew-scattering is dominant. Therefore, $\rho_H$ can be rewritten as $\rho_H = R_O B + c\rho M$. On the other hand, magnetization of the ferromagnet in the paramagnetic state ($T \geq T_C$) is expressed as $M = \chi H = \chi B/\mu_0$, where magnetic susceptibility $\chi$ follows the Curie Weiss law, $\chi = C/(T-T_C)$, and $C$ is the Curie constant. Eventually, by taking into account the general Curie Weiss equation, the Hall coefficient $R_H$ is expressed as $R_H = \rho_H/B = R_O + c\rho\, C/\mu_0(T-T_C)$. The fitting using this equation to the experimental $R_H$ -$T$ trace yields Curie temperature $T_C$ and the carrier concentration $p$ (= $1/eR_O$; Here, $e$ is the charge of a hole).

It is notable that the Curie constant $C$ of diluted magnetic semiconductors is theoretically expressed as $C = (4/a_0^3)(p_{eff}^2\mu_B^2 x)/(3k_B)$ [13], where $\mu_B$ is the Bohr magneton (1.16542×10$^{-29}$ Wb.m), $k_B$ is the Boltzman constant (1.3805×10$^{-23}$ J/T), $a_0$ is the lattice constant (in our case, $a_0$ is set at 5.65Å, the value for GaAs), $x$ is the Mn content (in our case, $x$ is set at 0.15 for 0.3ML Mn since we defined the local Mn content $x = \theta_{Mn}$ /2ML at $\theta_{Mn}$ < 1ML), $p_{eff}$ is the effective magnetic moment of a Mn ion, and $p_{eff}$ is defined as $p_{eff} = g\,[S_{Mn}(S_{Mn}+1)]^{1/2}$ (Here, $g$ and $S_{Mn}$ are $g$-factor and total spin of Mn). In the (GaMn)As random alloy [12], $g$ = 2, and $S_{Mn}$ = 5/2. We used the same values in our analysis of the Mn δ-doped GaAs layers. The above general equation of a bulk magnetic material can be modified to $R_{H\text{-}sheet} = R_{O\text{-}sheet} + cR_{sheet}C/\mu_0(T-T_C)$ for a 2-dimensional system, where $R_{H\text{-}sheet}$ is the sheet Hall coefficient, $R_{O\text{-}sheet}$ is the ordinary sheet Hall coefficient and $R_{sheet}$ is the sheet resistance of the sample. From the fitting using the modified equation to $R_{H\text{-}sheet}$ measured at $T \geq T_C$, one can estimate $R_{O\text{-}sheet}$ ($R_{O\text{-}sheet} = e/p_{sheet}$), thus the sheet hole concentration $p_{sheet}$, and the Curie temperature $T_C$.

Figure 5 (a) shows the temperature dependence of the sheet resistance $R_{sheet}$ and sheet Hall coefficient $R_{H\text{-}sheet}$ of the as grown 0.3ML Mn δ-doped GaAs ($T_s$ = 300°C) / Be-doped $p$-AlGaAs heterostructure. In the paramagnetic state of the sample ($T \geq T_C$), the experimental value of the $R_{H\text{-}sheet}$ (in the Ohm/Tesla unit) is expressed as the slope of the linear relation of Hall resistance versus magnetic field $B$. In other word, $R_{H\text{-}sheet}$ is the Hall resistance at $B$ = 1 Tesla. Figure 5 (a) also plots the Curie Weiss fitting to the experimental $R_{H\text{-}sheet} - T$ data measured at different temperatures ($T \geq T_C$), and also shows temperature dependent curves of magnetic susceptibility $\chi$ (= $C/(T-T_C)$), sheet ordinary

Hall coefficient $R_{\text{O-sheet}}$, and $cR_{\text{sheet}}\chi/\mu_0$, needed for the Curie Weiss fitting. In the top trace of measured sheet resistance $R_{\text{sheet}}$, a semiconductor-type behavior is observed. Near the sudden increase of resistance at low temperature ($T$ < 110K), there appeared a local maximum at around 112K, which might be at the position of its Curie temperature. From the Curie Weiss fittings, it was estimated that the Curie temperature $T_C$ was 112K, and the sheet hole concentration $p_{\text{sheet}}$ was $1.4\times10^{12}$ cm$^{-2}$. The proportional constant $c$ was taken to be 2.2. The estimated Curie temperature $T_C$ of 112K is in good agreement with the local maximum of $R_{\text{sheet}}$ at 112K and also with the temperature dependence of the remanence of the Hall hysteresis loops at $T \leq T_C$. The Curie temperature $T_C$, estimated in the same way, of samples annealed at $T_a$ = 280, 300 and 320°C were 120K, 172K, and 128K, respectively, which were in good agreement with the temperature

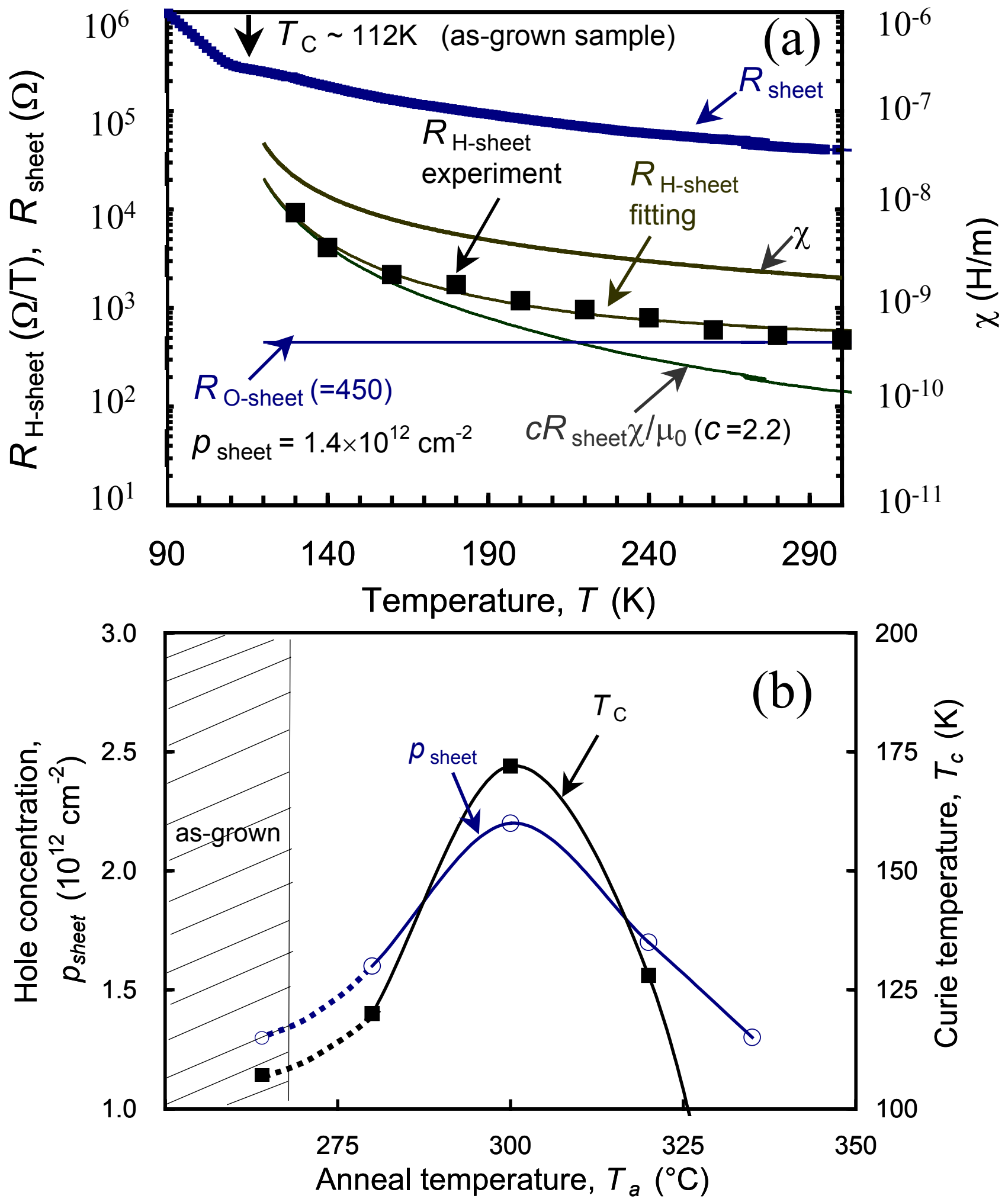

**Figure 5**. (a) Temperature dependence of the sheet resistance $R_{\text{sheet}}$ and sheet Hall coefficient $R_{\text{H-sheet}}$ (in the Ohm/Tesla unit) of the as-grown 0.3ML Mn δ-doped GaAs ($T_s$ = 300°C) / Be-doped $p$-AlGaAs heterostructure. From the Curie Weiss fitting to $R_{\text{H-sheet}}$, the estimated sheet ordinary Hall coefficient $R_{\text{O-sheet}}$ and magnetic susceptibility $\chi$ yield sheet hole concentration $p_{\text{sheet}}$ = $1.4\times10^{12}$ cm$^{-2}$ and Curie temperature $T_C$ = 112K. Here, the sheet Hall coefficient $R_{\text{H-sheet}}$ is equal to $R_{\text{O-sheet}} + cR_{\text{sheet}}\chi/\mu_0$ in the above traces, where $R_S = cR_{\text{sheet}}$ and $\chi = C/(T\text{-}T_C)$. (b) $T_C$ and $p_{\text{sheet}}$ as a function of annealing temperature $T_a$. The as-grown results are connected with dotted lines.

dependence of the remanence of the Hall hysteresis loops at $T \leq T_C$ (see Fig. 4).

In order to explain the remarkable enhancement of $T_C$ due to the LT- annealing, the estimated $T_C$ and $p_{sheet}$ using Curie Weiss plot is shown as a function of annealing temperature $T_a$ in Fig. 5 (b). It is apparent that the decrease in electrical compensation by As-rish donors resulted in the increase in the hole concentration and magnetically active Mn dopants, and thus enhanced 2DHG-mediated ferromagnetic ordering among local Mn spins and increased its $T_C$. However, at $T_a$ = 335°C, diffusion of the Mn dopants might have resulted in the destruction of ferromagnetic ordering.

## 5. Conclusions

In conclusion, we have prepared Mn δ-doped GaAs / Be-doped AlGaAs heterostructures with focus on the *p*-type selective doping effect on the Mn δ-doped GaAs layers. Intentionally provided holes from the Be doped AlGaAs layer to the Mn δ-doped GaAs resulted in ferromagnetic ordering with the $T_C$ of 70K (grown at $T_s$ = 400°C). By lowering the growth temperature of Mn δ-doped GaAs layer to 300°C, the Mn dopant profile became more abrupt with high peak, leading to the higher Curie temperature of 112K. Furthermore, LT-annealing treatment significantly enhanced the magnetic property with the highest Curie temperature $T_C$ of 172K. This $T_C$ value is the highest among the reported III-V (InAs and GaAs) based magnetic semiconductors.